# Electrical switching of spin-polarized light-emitting diodes based on a 2D CrI$_3$/hBN/WSe$_2$ heterostructure

Jianchen Dang[1], Tongyao Wu[1], Shuohua Yan[2,3], Kenji Watanabe[4], Takashi Taniguchi[5], Hechang Lei[2,3], Xiao-Xiao Zhang[1*]

[1]*Department of Physics, University of Florida, Gainesville, FL, United States*

[2]*Department of Physics and Beijing Key Laboratory of Opto-electronic Functional Materials & Micro-nano Devices, Renmin University of China, Beijing 100872, China*

[3]*Key Laboratory of Quantum State Construction and Manipulation (Ministry of Education), Renmin University of China, Beijing 100872, China*

[4]*Research Center for Functional Materials, National Institute for Materials Science, 1-1 Namiki, Tsukuba, Japan*

[5]*International Center for Materials Nanoarchitectonics, National Institute for Materials Science, 1-1 Namiki, Tsukuba, Japan*

\*Corresponding Author: xxzhang@ufl.edu

## Abstract

Spin-polarized light-emitting diodes (spin-LEDs) convert the electronic spin information to photon circular polarization, offering potential applications including spin amplification, optical communications, and advanced imaging. The conventional control of the emitted light's circular polarization requires a change in the external magnetic field, limiting the operation conditions of spin-LEDs. Here, we demonstrate an atomically thin spin-LED device based on a heterostructure of a monolayer WSe$_2$ and a few-layer antiferromagnetic CrI$_3$, separated by a thin hBN tunneling barrier. The CrI$_3$ and hBN layers polarize the spin of the injected carriers into the WSe$_2$. With the valley optical selection rule in the monolayer WSe$_2$, the electroluminescence exhibits a high degree of circular polarization that follows the CrI$_3$ magnetic states. Importantly, we show an efficient electrical tuning, including a sign reversal, of the electroluminescent circular polarization by applying an electrostatic field due to the electrical tunability of the few-layer CrI$_3$ magnetization. Our results establish a new platform to achieve on-demand operation of nanoscale spin-LED and electrical control of helicity for device applications.

## Main

The spin states of materials are the building blocks of modern information technology. Most spintronics devices achieve the control and detection of electronic spin based on electrical currents. In a spin-polarized light-emitting diode (spin-LED), the injection of spin-polarized carriers results in circularly polarized electroluminescence (EL), interfacing optoelectronics and photonics with spintronics[1]. Spin-LEDs have been demonstrated using GaAs-based ferromagnet/semiconductor structures[2-4], organic semiconductors like chiral molecules[5] and hybrid perovskites[6], and two-dimensional (2D) layered heterostructures[7,8], with some showing capabilities of room-temperature operation. However, controlling the degree of polarization in the EL signal for these spin-LED devices often requires a change in the temperature, magnetic field, or chemical composition. Efficient electrical control of the EL polarization will enable low-power and high-speed applications in spin-optoelectronics[1], information processing[9], and ellipsometry-based tomography[10,11].

The recent advances in 2D layered materials open up new possibilities for optoelectronics and spintronics device designs that are more flexible and tunable. In a monolayer semiconducting transition metal dichalcogenide (TMD), the valley-spin coupling and the valley-dependent optical selection rule ensure that we can selectively determine the circular polarization of the emitted light based on the carriers' and excitons' valley and spin occupation[12]. In addition, 2D TMDs have remarkable optical properties due to their large excitonic interactions, and their optoelectronic prototypes like photodetector and light-emitting diodes have been demonstrated[13]. The van der Waals magnetic crystals have been shown to maintain magnetic ordering down to monolayer or few-layer limit, enabling the construction of nanoscale spintronic devices and easy integration with other 2D systems[14,15]. In particular, $CrI_3$ crystals are A-type antiferromagnets at the few-layer limit, with the spin easy axis along the out-of-plane direction. For an even or odd layer number, the corresponding magnetic states in $CrI_3$ are overall ferromagnetic or antiferromagnetic. The magnetic interactions, including the magnetic resonance frequency, in few-layer $CrI_3$ can be efficiently tuned by either an out-of-plane electric field or doping density, providing a unique opportunity to develop electrically tunable magnetic devices.

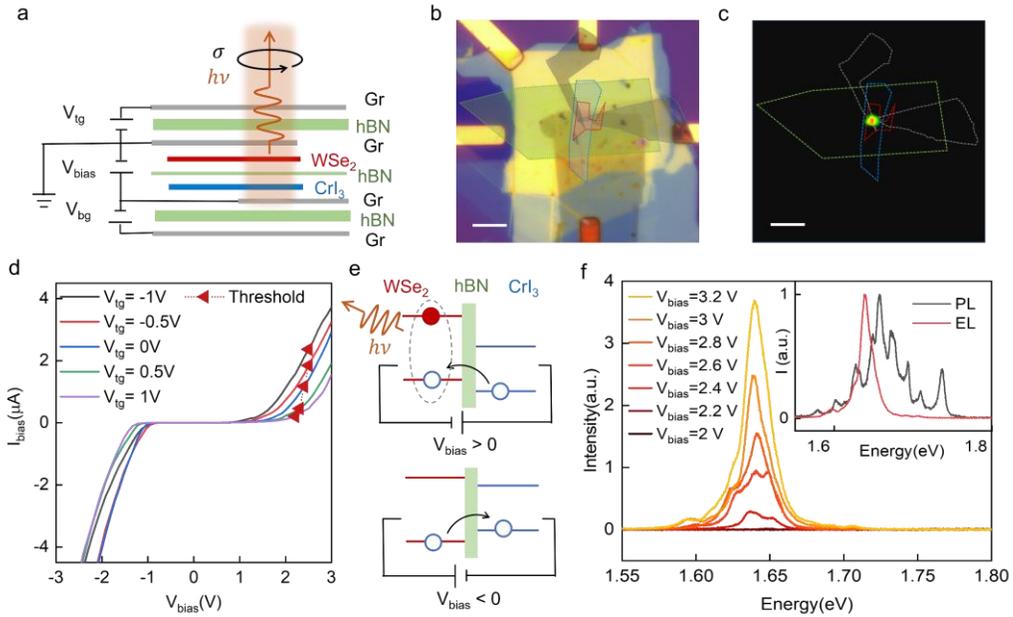

**Figure 1. Electroluminescence based on WSe$_2$/hBN/CrI$_3$ heterostructure. a,** Schematic of the device structure. The back gate $V_{bg}$ and top gate $V_{tg}$ is used to control the doping of CrI$_3$ and WSe$_2$, and the bias voltage is applied across the vertical junction going through the hBN tunneling barrier. **b,** The whitelight microscope image of a monolayer WSe$_2$/hBN/bilayer CrI$_3$ device. The scale bar is 10 μm. The monolayer WSe$_2$ is outlined by red dashed lines, blue dashed lines outline CrI$_3$, the thin hBN (~2.5 nm) tunneling layer is denoted by the green dashed lines, and the source and drain contacts (from graphite stripes) are denoted by the gray lines. **c,** Spatially-resolved EL image of the device. The notation of dashed lines is the same as **b**. The EL signal comes from the overlapped area of the source and drain contacts. **d,** I–V characteristics of the heterostructure at different top gate voltages while keeping $V_{bg} = 0V$. The red triangles denote the extracted threshold point for EL generation. **e,** Band alignment schematics of the heterostructure under different bias voltages for the scenarios of EL generation with a positive $V_{bias}$ (upper panel, WSe$_2$ n-doped) and no EL generation with a negative $V_{bias}$ (lower panel). **f,** EL spectra at different $V_{bias}$ with $V_{tg}$ =1V and $V_{bg}$=0V. The inset shows the comparison of the normalized PL and EL spectra. The PL spectra was taken with a 633 nm continuous wave laser (10 μW). The EL spectrum was measured under $V_{bias}$=2.5V, $I_{bias}$=0.49 μA with $V_{tg}$=1V.

Here, we fabricated 2D LED structures based on monolayer $WSe_2$/hBN/few-layer $CrI_3$, where the doping of $WSe_2$ and $CrI_3$ can be individually controlled by separate gating electrodes (Fig. 1a). The back gate $V_{bg}$ and top gate $V_{tg}$ is used to control the doping of $CrI_3$ and $WSe_2$, respectively. The $CrI_3$ will serve as a spin-polarizing layer for carrier injection. When at appropriate bias voltages and doping levels, p-type carriers are injected through the $CrI_3$/hBN into the n-doped monolayer $WSe_2$, which gives rise to $WSe_2$ electroluminescence signals. In order to maintain the spin polarization during the carrier injection, a thin hBN tunneling barrier (the thickness is confirmed with the atomic force microscope measurement, as shown in Supplementary Fig. S1) is used to overcome the impedance mismatch[16] and also to avoid the Schottky barrier at the interfaces. Two graphite stripes contact the $WSe_2$ and $CrI_3$ separately and serve as the source and drain contacts. Figure 1b shows the microscope image of a device with a bilayer $CrI_3$. The Methods section provides detailed information on material preparation and heterostructure fabrication.

## EL characterization

When applying a bias voltage between the $CrI_3$ and $WSe_2$, the tunneling current $I$ starts to flow for both positive and negative applied bias voltages, as shown in Fig. 1d. The EL from the $WSe_2$ can only be observed for the positive bias region, and the onset EL threshold current decreases with an increasing top gate voltage, and therefore at a higher n-type doping density in $WSe_2$ (The back gate $V_{bg}$ for the $CrI_3$ doping is kept at zero for this part of the experiment). Combined with the expected type-II band alignment between the $WSe_2$ and $CrI_3$[17] and the measured doping level shift from the $WSe_2$ photoluminescence (PL) (see Supplementary Fig. S2), the *I-V* characteristic and the EL bias dependence can be understood by considering the type-II to type-I band alignment transition when applying a negative bias voltage, as depicted in Fig. 1e. With a positive bias voltage, p-type carriers flow from the $CrI_3$/hBN layer into $WSe_2$. They can recombine with the n-type carriers in $WSe_2$ and generate EL signals. On the other hand, with a negative bias voltage, there is no EL signal in $WSe_2$ when tuning the $WSe_2$ doping level from p- to n-type doping. It thus indicates that there is no carrier injection into $WSe_2$ with a negative $V_{bias}$ (when there is significant tunneling current). This is likely due to the shift in $CrI_3$ bands under the bias voltage, which converts the type-II to a type-I band alignment. The p-type carriers can flow from $WSe_2$ to $CrI_3$ layers and give rise to tunneling currents without EL generation.

From the spatially-resolved EL imaging in Fig. 1c, the EL generation is most efficient at the overlapping region of the source and drain graphite strip, where the tunneling current goes through the vertical heterostructure without further diffusion or drift. Figure 1f shows the evolution of the EL signals at different bias voltages and tunneling currents when $V_{tg}$=1V and $V_{bg}$=0V. The top gate dependence of the EL spectra is plotted in

Supplementary Fig. S2a. The EL signal quenches when the WSe$_2$ is tuned to p-doped, which can be deduced from the gate-dependent PL in Fig. S2b, consistent with the expected EL generation process in Fig. 1e. The *I-V* curves and EL spectra at different back gate voltages were summarized in Supplementary Fig. S2, which do not show significant back gate dependence. The comparison of the EL and PL (at $V_{tg} = 0$) spectra is plotted in the inset of Fig. 1f. The assignment of exciton species for PL spectra is shown in Supplementary Fig. S3. The EL emission peak is redshifted with no obvious defect-related peaks, which may be attributed to the additional carrier screening and the charge-related defect states being filled up. A more detailed comparison and analysis of the EL exciton contributions can be found in Supplementary Information Fig. S3.

## Spin-dependent circularly polarized EL

We measured the magnetic field-dependent EL to reveal the spin sensitivity of this structure. Under an out-of-plane magnetic field, the few-layer CrI$_3$ will go through layer-dependent spin-flip transitions. When the CrI$_3$ layer is spin-polarized, the CrI$_3$/hBN will serve as a spin-filtering layer for the injected holes into the WSe$_2$ layer. The valley-spin coupling and valley-dependent optical selection rule in WSe$_2$ subsequentially generate circularly polarized light emission based on the injected hole spin polarization (Fig. 2a). The large spin-orbit coupling splitting in the WSe$_2$ valence bands and long valley lifetime of the holes further facilitate the generation of the circularly polarized EL. Figure 2b shows the oppositely circularly polarized EL spectra at opposite magnetic fields from a bilayer CrI$_3$/hBN/WSe$_2$ device (± 1.8T is a fully polarizing field for bilayer CrI$_3$). As shown in Fig. 2 c&d, we measured and compared the magnetic state switching of the CrI$_3$ layer through the reflective magnetic circular dichroism (RMCD) and the EL light helicity switching with different magnetic field ramping directions. Here, the EL polarization is characterized by the helicity as defined by $(I_{\sigma+} - I_{\sigma-})/(I_{\sigma+} + I_{\sigma-})$. The RMCD shows the spin-flip transition that corresponds to the layer-dependent spin switching in bilayer CrI$_3$[18], as indicated by the schematic plot in Fig. 2c. The EL helicity follows the RMCD magnetic field dependence and shows a jump to ~ ±10% at the spin-flip fields. This phenomenon was further confirmed by measuring three additional devices with bilayer CrI$_3$. In all cases, the EL helicity followed the RMCD traces (see Supplementary Fig. S4). As the temperature was increased to be close to the Neel temperature of CrI$_3$ (~ 45K), the EL helicity also dropped to zero (see Supplementary Fig. S5).

Depending on the layer number of the CrI$_3$, we can further tune the EL helicity field dependence. A trilayer CrI$_3$ is ferromagnetic at zero fields and goes through layer dependent spin-flip transitions with an increasing out-of-plane magnetic field (Fig. 2e). The corresponding EL helicity of a trilayer CrI$_3$/hBN/WSe$_2$ device also shows zero field helicity and spin-flip fields consistent with the RMCD signals, as shown in Fig. 2f. The measured EL helicity varies across different devices, possibly due to variations in device quality. We discussed the variability of EL helicity and present data for multiple bilayer and trilayer CrI$_3$ devices in Supplementary Table 1. The maximum saturation polarization obtained was ~ 40%, as shown in this trilayer device. This helicity in EL is intrinsically limited mainly by the exciton depolarization, which arises from the efficient intervalley exciton exchange interactions[19]. Supplementary Fig. S7 is the measured circular polarization of neutral and charged excitons in PL with a near-resonant valley-polarized optical excitation, with trion states showing a maximum of ~ 37% circular polarization, close to the highest circular polarization observed in EL devices. We therefore infer the spin filter efficiency was close to unity with these 2D magnetic tunneling junctions in the device with observed maximum EL helicity, and the helicity was mostly constrained by the intervalley exciton depolarization.

Notably, the EL helicity is determined by the overall magnetization of the $CrI_3$, instead of the topmost layer adjacent to the $WSe_2$ layer, which is distinctly different from the previously reported circularly polarized PL quenching in $CrI_3/WSe_2$[20]. In a $CrI_3/WSe_2$ structure, spin-dependent charge transfer is mostly determined by the adjacent $CrI_3$ layer spin polarization. In comparison, the hBN barrier here (Fig. 1a) ensures that the tunneling carriers' spin is set by the overall magnetization of the CrI$_3$ layer. The EL helicity does not show observable dependence on the WSe$_2$ layer doping level (Fig. 3a) and the applied bias voltage (Fig. 3b) within the EL generation ranges. Such consistent spin-polarized EL operation is consistent with the spin tunneling behavior, which is not sensitive to the relative shifts of Fermi levels across the heterostructure. To reveal the impact of the tunneling junction, we also measured EL signals with CrI$_3$/WSe$_2$ devices without the hBN tunneling barrier. While EL signals can still be observed, there is no obvious circular dichroism that depends on the CrI$_3$ magnetization (see Supplementary Fig. S6), which highlights the importance of tunneling barriers to reduce conductance mismatch and increase spin filter efficiency in 2D heterostructures.

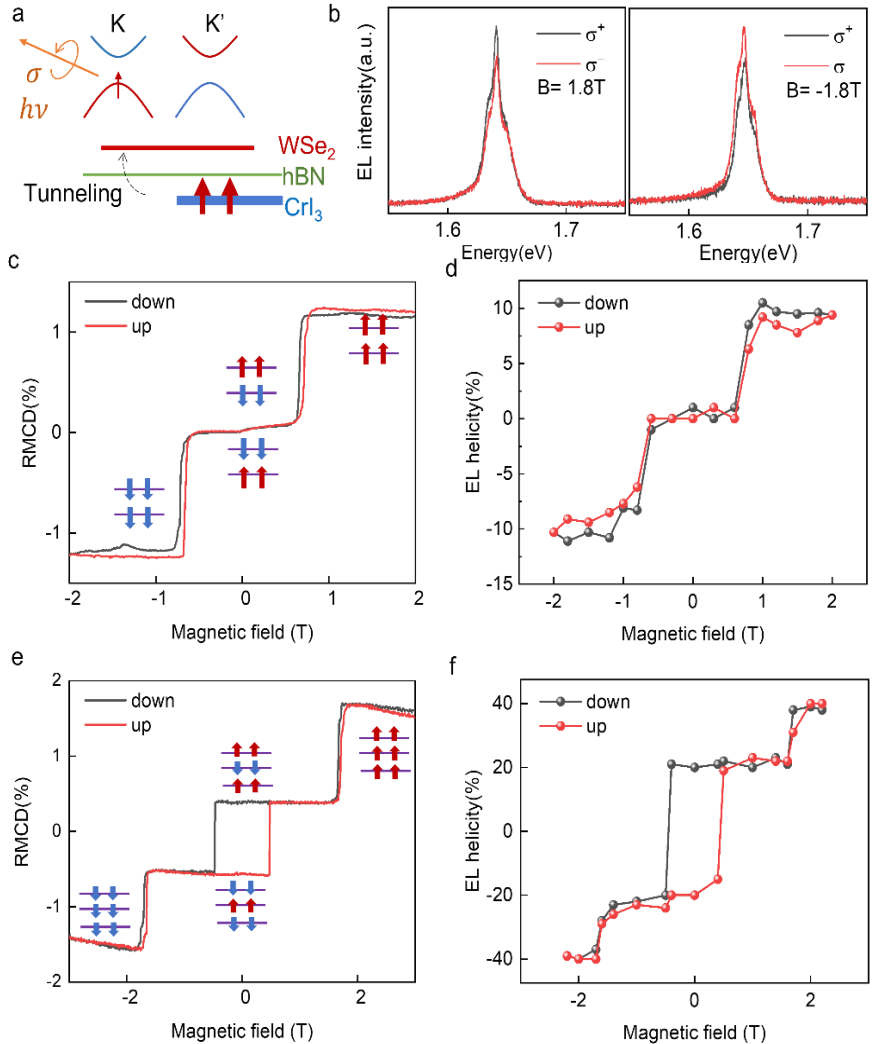

**Figure 2. Spin-dependent EL. a,** The spin dependence in EL signals originated from the spin-polarized carrier injection through $CrI_3$/hBN and the coupled spin and valley indexes in monolayer TMD. **b,** Polarization-resolved EL spectra under ±1.8 T out-of-plane magnetic fields, showing opposite helicity. **c,** The RMCD signals as a function of magnetic fields for a bilayer $CrI_3$/hBN/$WSe_2$ device. The spin switching for each layer is sketched. The up and down indicates the magnetic field sweeping direction. **d,** The corresponding extracted EL helicity as a function when sweeping the magnetic field. **e,** The RMCD signals and **f,** corresponding EL helicity of a trilayer $CrI_3$/hBN/$WSe_2$ device as a function of magnetic fields.

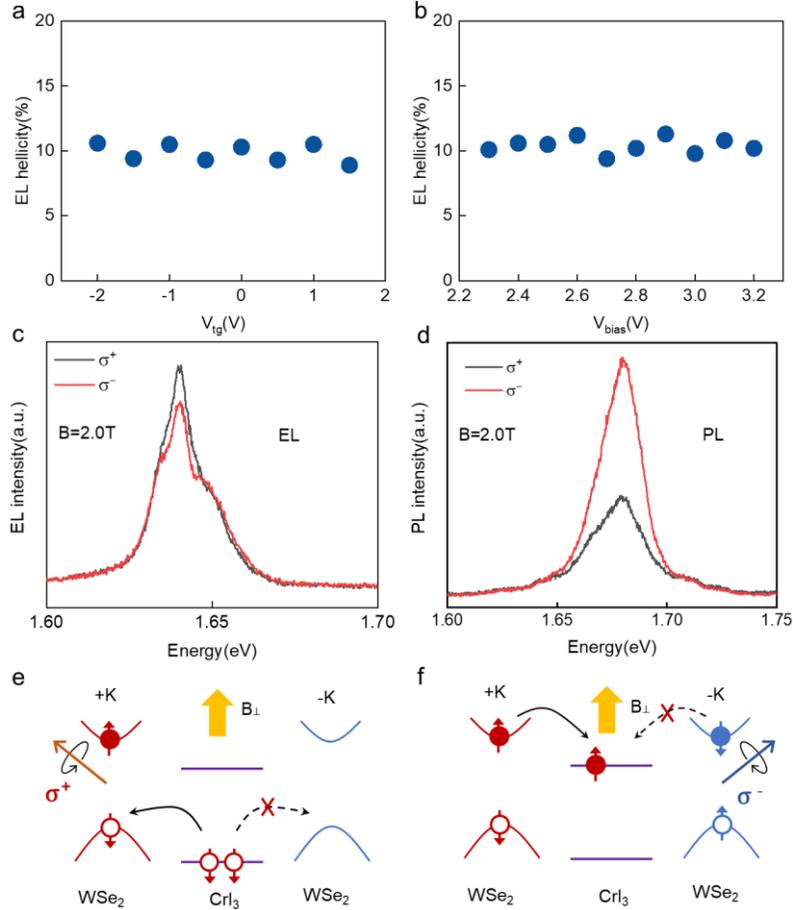

**Figure 3. Characterization of EL helicity. a,** The evolution of EL helicity (bilayer CrI$_3$/hBN/WSe$_2$) under different V$_{tg}$, measured under 2T out-of-plane magnetic field and with V$_{bias}$ = 2.5 V. **b,** EL helicity dependence when varying V$_{bias}$ (after reaching the EL threshold conditions), as measured with 2T out-of-plane field and V$_{tg}$ = 1V. **c & d** compares the circular polarization of the EL and PL signals at 2T magnetic field. The PL was measured at a WSe$_2$/CrI$_3$ heterostructure region without hBN barrier, excited by linearly polarized 633nm laser. The EL spectra were measured in a CrI$_3$/hBN/WSe$_2$ device. EL and PL possess opposite circular polarization. **e,** The exemplary schematics illustrate the spin-polarized carrier injection process, where the spin polarization results in K valley and σ+ EL emission. **f,** Under the same CrI$_3$ spin alignment, the interlayer charge transfer favors the quenching of K valley electrons and therefore results in a higher -K exciton population and σ- PL emission.

To further illustrate the differences in the tunneling spin injection and spin-dependent charge transfer, we further compared the helicity of PL in a $CrI_3/WSe_2$ heterostructure and EL spectra in a $CrI_3/hBN/WSe_2$ device. Figure 3d shows the circularly polarized PL taken with a linearly polarized excitation in $CrI_3/WSe_2$ under a 2T magnetic field. The enhanced PL polarization is caused by charge transfer, consistent with previous work[17,20]. In comparison, the EL polarization (Fig. 3c) is oppositely polarized. This is consistent with the expectation of the band alignments and transfer processes. The EL polarization is determined and aligned by the $CrI_3$ spin direction. As shown in Fig. 3e, injected spin-polarized carriers will reside in, e.g., the K valley because of the valley-spin coupling in monolayer TMD and gives rise to σ+ emission. On the other hand, the PL polarization (on heterostructure area) arises from the spin-dependent charge transfer[20,21], which quenches the valley/spin-polarized exciton with carriers' spin aligned with the $CrI_3$ spin orientation. In the depicted scenario in Fig. 3f, under the same $CrI_3$ spin alignment as Fig. 3e, the K valley exciton will be quenched due to electron interlayer transfer, giving rise to an overall σ- polarization in light emission.

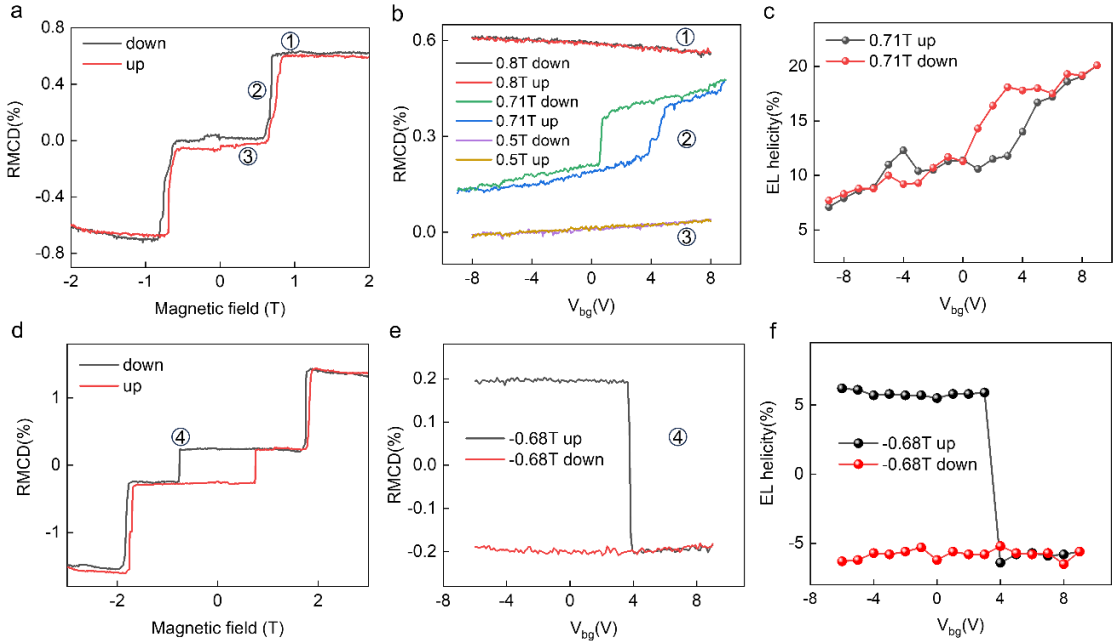

**Figure 4. Electrical switching of EL helicity. a,** RMCD of bilayer $CrI_3$ as a function of magnetic field under zero back gate voltage at 4 K. **b,** Back gate voltage control of RMCD of bilayer $CrI_3$ under different $V_{bg}$ sweeping cycles. The sample was prepared in the "up" state by a magnetic field at 2T and then biased at 0.8, 0.71 and 0.5 T corresponding to traces 1 to 3 in **a** and **b**, respectively. The up and down in **b** indicates the sweeping direction of $V_{bg}$. **c,** The repeatable switching of EL helicity when varying the back gate voltage at a fixed 0.71T, which corresponds to trace 2 position in **a** and **b**. **d,** RMCD of trilayer $CrI_3$ as a function of magnetic field under zero back gate voltage at 4 K. **e,** Back gate voltage control of RMCD of trilayer $CrI_3$. The sample was prepared by a magnetic field at 2T and then biased at -0.68T (trace 4). **f,** The EL helicity of the corresponding trilayer $CrI_3$ device when varying $V_{bg}$ at a fixed -0.68T (trace 4 position).

## Electrical switching of EL helicity

Efficient electrical tuning of magnetic interactions has been demonstrated in few-layer CrI$_3$ in prior studies[22-25]. Here, we utilize the electrical tunability of CrI$_3$ to control the spin-dependent EL signals. To this end, we use the back gate to electrostatically tune the doping level in CrI$_3$ (Fig. 1a) while observing the EL helicity switching. When varying the doping in CrI$_3$, the spin-flip transition field can show significant shifting[22,24] and gives rise to spin switching, and therefore EL helicity switching, at certain fixed magnetic fields. In Figure 4a, we prepared a bilayer device in the "up" state by applying a magnetic field of 2 T. Subsequently, we swept the back gate voltage while maintaining the system at 0.8, 0.71, and 0.5 T, corresponding to traces 1 to 3 in Fig. 4b, respectively. When at magnetic fields (0.8T and 0.5T) away from the spin-flip field, the magnetization remains in ferromagnetic and antiferromagnetic states, respectively, as shown in the RMCD measurements. Near the spin-flip field (trace 2), a repeatable switching between the ferromagnetic and antiferromagnetic states can be achieved, consistent with previous reports[22,24]. The measured EL helicity shows corresponding repeatable switching between 7% and 21% (Fig. 4c), with switching gate voltage hysteresis similar to that observed in RMCD. We note that the incomplete AFM to FM switching here is due to the limited back gate voltages applied in this device. Alternatively, we also examined the switching capability with a trilayer CrI$_3$ device (Fig. 4d). The device was initially prepared at 2 T and then subjected to a fixed magnetic field just below its coercive force at -0.68 T. As the gate voltage scanned from negative to positive values, it induces a sign switch in the magnetism of CrI$_3$ (Fig. 4e) because of the decrease in the spin-flip field. Notably, this is a one-time-only switching event, as the device remains in the negative sign state afterward due to it being the low-energy state under a negative magnetic field. Due to the spin reversal in CrI$_3$, it thus gives rise to a sign reversal in the EL helicity that is triggered by electrical signals, as shown in Fig. 4f. In addition, we investigated the repeatable switching behavior near the spin-flip transition field of 1.73T, as shown in Supplementary Fig. S8.

## Conclusions

In conclusion, we showed robust spin-LED device operation composed of CrI$_3$/hBN/WSe$_2$ van der Waals heterostructures, where the spin-polarized carriers tunneled through the CrI$_3$/hBN layer and resulted in valley polarized and circularly polarized light emission. A close-to-unity spin transfer efficiency was achieved with our tunneling contacts. Importantly, we demonstrate an effective modulation and control of EL helicity through electrical signals due to the electrical tunability of magnetization in CrI$_3$. Our results provide a novel approach to having on-demand, electrically tunable helicity in 2D spin-LED, opening up new directions to combine optoelectronics, spintronics, valleytronics, and advanced imaging.

## Methods

**Crystal growth**

CrI$_3$ single crystals were grown by the chemical vapor transport method. Chromium powder (99.99% purity) and iodine flakes (99.999%) in a 1:3 molar ratio are put into a silicon tube with a length of 200 mm and an inner diameter of 14 mm. The tube was pumped down to 0.01 Pa and sealed under vacuum, and then placed in a two-zone horizontal tube furnace. The two growth zones are raised slowly to 903 and 823 K for two days and are then held there for another seven days. Shiny, black, plate-like crystals with lateral dimensions of up to several millimeters can be obtained from the growth. In order to avoid degradation, the $CrI_3$ crystals are stored in an inert-gas glovebox.

**Device fabrication**
The few layer graphite, hBN, bilayer/trilayer $CrI_3$ and monolayer $WSe_2$ were first mechanically exfoliated from bulk crystals and identified by their color contrast under an optical microscope. The heterostructure was built by using the dry transfer technique with a PC stamp[26] and released onto a substrate with pre-patterned gold electrodes. The transfer steps were performed in a nitrogen-filled glove box.

**Optoelectronic measurements.**
The devices were mounted onto a 3D piezoelectric stage in an optical cryostat (attoDry1000) with a base temperature of 4 K. The cryostat was equipped with a superconducting solenoid magnet, which can supply a magnetic field from -9 T to 9 T. For EL and PL measurements, the emission was collected by an objective lens with a numerical aperture of 0.82 and detected by a grating spectrometer and CCD (Princeton Instruments SpectraPro HRS300 + PIXIS). The polarization of the emission was measured by using a $\lambda/4$ plate followed by a polarizer. For PL measurement, the sample was excited by a 633-nm continuous wave laser with a focal spot diameter ~1 μm.

**RMCD measurements**
For RMCD measurements, the 633-nm continuous wave laser was used. The laser was modulated at 50 kHz between the left and right circular polarization using a photoelastic modulator (Hinds PEM). The reflected light was focused onto a photodiode. The RMCD was determined as the ratio of the a.c. component of the photodiode signal measured by a lock-in amplifier at the polarization modulation frequency and the d.c. component of the photodiode signal measured by an oscilloscope.

## Data availability
The data that support the findings of this study are available within the paper and its Supplementary Information. Additional data are available from the corresponding authors upon request.

## Acknowledgment
X-X.Z. acknowledge the support from the Department of Energy (DOE) award DE-SC0022983. This work was partly conducted at the Research Service Centers of the Herbert Wertheim College of Engineering at the University of Florida. H.C.L. was supported by Beijing Natural Science Foundation (Grant No. Z200005), National Key R&D Program of China (Grants Nos. 2022YFA1403800, 2023YFA1406500), and National Natural Science Foundation of China (Grants Nos. 12274459).

## Author contributions
X-X.Z. and J. D. designed the study. J.D. and T.W. fabricated the device and performed the measurements. K.W. and T.T. grew the bulk hBN crystals. S.Y. and H.C.L. grew the bulk CrI$_3$ crystals. X-X.Z. and J.D. wrote the manuscript. All authors discussed the results and commented on the manuscript.

## Competing Interests
The authors declare no competing interests.